\documentclass[a4paper]{jpconf}
\usepackage{graphicx}

\begin{document}

\graphicspath{{Plots/}}

\title{Experimental systematic uncertainties (and object
  reconstruction) on top physics, their correlations, comparison ATLAS
vs CMS (vs Tevatron) and common agreements}

\author{M.J. Costa on behalf of the ATLAS, CDF, CMS and D0 Collaborations}

\address{Instituto de F\'\i{}sica Corpuscular, University of Valencia and CSIC, C/Catedr\'atico
  Jos\'e Beltr\'an, 2, E-46980 Paterna, Spain}

\ead{Maria.Jose.Costa@ific.uv.es}

\begin{abstract}
The experimental systematic uncertainties associated to the
reconstruction and calibration of the objects appearing in top quark
final states at the LHC and Tevatron are discussed. The strategies followed in the ATLAS and
CMS experiments are compared in detail for the cases of the jet energy
scale and $b$-tagging calibrations, where a categorisation of the
associated uncertainty sources as well as the corresponding correlations across
experiments has been proposed. The estimate of the non-prompt and fake
lepton background to the top quark leptonic channels is also discussed. 
\end{abstract}

\section{Introduction}

The precision of several top quark physics measurements is being
improved by combining the results obtained by different experiments of
the LHC and Tevatron accelerators. As an example, an LHC
\cite{LHCTopMassCombination}, Tevatron
\cite{TevatronTopMassCombination} and even a world average
\cite{WorldTopMassCombination} combination of the top quark mass have
recently become available. When performing these combinations it is
essential to understand the proper categorisation of systematic
uncertainties as well as the correlations across experiments for each
category. This includes the experimental uncertainties associated to
the reconstruction and calibration of the objects appearing in the top
quark final state under consideration: jets as well as $b$-jets and
for the leptonic decays channels also leptons and missing transverse
energy.

A detailed study has recently been done at the LHC for the jet energy
scale and $b$-tagging calibration uncertainties, resulting in a
proposal for the treatment of these uncertainties when performing
combinations of measurements provided by the ATLAS \cite{ATLAS} and
CMS \cite{CMS} experiments. These uncertainties are the dominant experimental systematic
uncertainties in most top quark measurements and will be therefore
discussed in more detail in this note.

At the LHC experiments, recent significant improvements on alignment, calibrations and material
description in simulations have led to a better understanding of
leptons
\cite{ATLASelectrons1,ATLASelectrons2,CMSelectrons,ATLASmuons,CMSmuons}. Uncertainties
on the lepton efficiency data/MC scale factors, scale and resolution
calibrations are considered in top quark measurements. For what
concerns the missing transverse energy performance, the main focus has
been put in developing techniques to cope with the high pile-up
activity in the 2012 data \cite{ATLASMETPileup,CMSMETPileup}. The
related uncertainties come from the scale and resolution uncertainties
of the reconstructed objects used in the missing transverse energy
reconstruction, as well as from the description of the energy not
associated to such objects.

In the top quark leptonic final states, after the top quark event selection which
includes the identification of isolated leptons, there is still a
contribution from non-prompt leptons and other particles (as jets)
which are mis-identified as leptons. The most commonly used data
driven techniques to estimate the contribution of this background have
recently been documented in a dedicated note by the ATLAS
Collaboration \cite{ATLASfakes} and will also be summarised here.

\section{Jets}

A different jet reconstruction algorithm is used in each experiment:
Cone algorithms are used at CDF and D0 experiments
\cite{JetAlgCDF,JetAlgD0}, including midpoints in the list of seeds
for the case of D0, and with different cone sizes of 0.4 (0.5) at CDF
(D0). At the LHC, the more recent anti-$k_t$ algorithm \cite{antikt}
is used instead with a cone size of 0.4 for the case of ATLAS and 0.5
for CMS.

Jets are reconstructed from calorimeter energy deposits in all
experiments except for the case of CMS where particle flow
candidates \cite{PFCMS} are used as input (excluding those charged
hadrons associated with a pile-up vertex). For the purpose of
calibration, the jet algorithm is also run on simulated stable
particles, excluding those that do not leave a visible energy in the
detector (neutrinos), and for the case of ATLAS excluding also muons. 

Jet quality criteria as well as pile-up rejection cuts are applied
within the different top quark analyses on the calibrated
jets. Uncertainties associated to the jet energy scale calibration,
resolution and efficiency are considered in these analyses.

The jet calibration at the LHC is designed in order to restore the jet
energy to that of jets from stable particles as described above
(so-called truth jets).  The procedure contains three main steps which
are common in ATLAS and CMS \cite{ATLASJES1,ATLASJES2,CMSJES}: 

\begin{itemize}

\item {\bf Pile-up correction:} The energy in the
  jet coming from pile-up is estimated and then subtracted. In the ATLAS experiment, for
  data collected in 2011 at $\sqrt{s}=7$~TeV, this offset is
  estimated based on Monte Carlo (MC) simulations, having the offset
  proportional to quantities related with both in-time and out-of-time
  pile-up. With the increase of the pile-up activity in 2012, an improved
  technique is used: the jet area correction method \cite{ATLASJES2} in which the pile-up
  contribution is estimated on an event by event basis, with an
  additional residual correction to obtain an average response
  insensitive to pile-up across the full $\eta$ range. CMS uses for
  all data periods the so-called hybrid jet area technique
  \cite{CMSJES}. The main differences with respect to the
  ATLAS jet area technique are that it takes care of not subtracting
  the underlying event contribution and that it only corrects for
  in-time pile-up, covering the smaller out-of-time effects with an
  uncertainty.   

\item {\bf Response correction:} This correction is based on simulation
  and corrects the energy of the reconstructed jets such that it is
  equal on average to the energy of the generated MC particle
  jets. The correction is computed for different bins in $p_{\rm T}$ and
  $\eta$ as the inverse of the jet response. In both ATLAS and CMS, it
  is estimated using isolated jets from an inclusive jet MC
  sample generated with Pythia. As mentioned before, the definition of
  truth jets is not exactly the same due to the fact that in ATLAS
  muons are not considered. However, this has been checked to have a
  negligible effect.

\item {\bf Residual in-situ correction:} In-situ techniques are used to
  correct for the remaining differences between data and MC in the jet
  response. This correction is only applied to data. Several
  techniques have been used exploiting the $p_{\rm T}$ balance between a
  well measured object (like a photon or a $Z$ boson) and the jet to
  be calibrated. A relative calibration is first performed to equalise the
  jet response in the forward region to that of the central region
  using events with only two jets at high $p_{\rm T}$. Central jets are then
  calibrated using $Z$+jets (with $Z$ decaying to $e^{+}e^{-}$ or
  $\mu^{+} \mu^{-}$) and $\gamma$+jets events in the so-called absolute calibration
  step. To extend the $p_{\rm T}$ range reached by these samples, ATLAS uses
  multi-jet events in which a system of now well calibrated low
  $p_{\rm T}$
  jets recoils against a high $p_{\rm T}$ jet. The assumed balanced between
  the reference object and the probe jet used in these in-situ
  techniques, can be affected by physics effects as additional
  radiation. The CMS strategy is always to extrapolate to the
  case of no parton radiation, while ATLAS accounts for this effect only in the
  uncertainty. The results obtained by the different techniques on the
  data/MC response ratios are then combined.

\end{itemize} 

Since the jet response depends on the jet flavour, an additional
flavour uncertainty is needed for topologies with different jet
flavour composition than those used to derive the calibration, as top
quark events. This uncertainty takes into account how well the flavour
response differences are known by comparing the predictions from
Pythia and Herwig++ MC generators.

\begin{figure}[h]
\begin{minipage}{16pc}
\includegraphics[width=15pc]{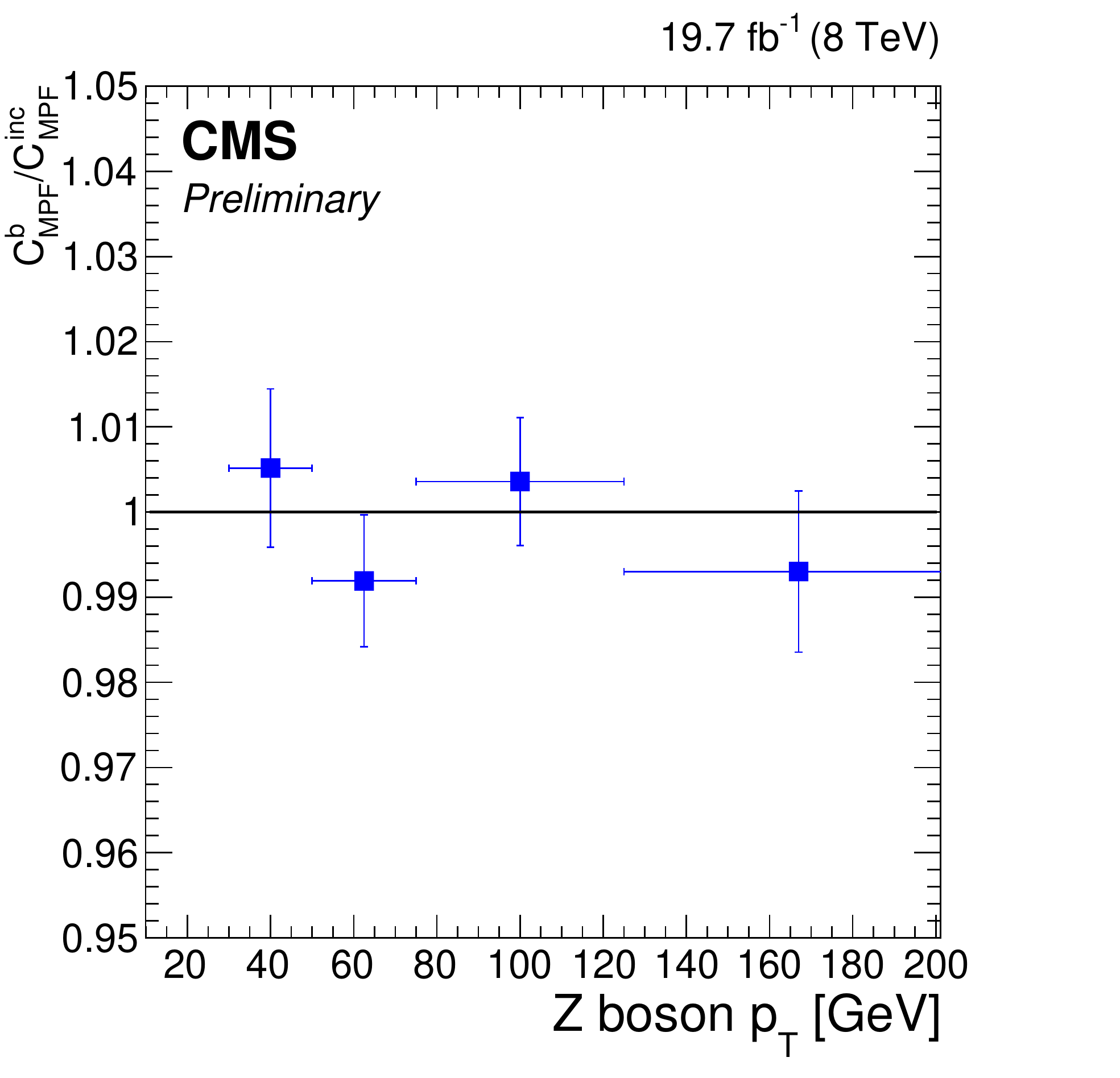}
\caption{\label{fig:bJetCorrection}Energy correction for $b$-jets as a function
  of the $Z$ boson $p_{{\rm T}}$ obtained as the ratio of the
  $b$-tagged and inclusive jet energy corrections in $Z+b$ events \cite{CMSbCorrection}.}
\end{minipage}\hspace{2pc}%
\begin{minipage}{16pc}
\includegraphics[width=15pc]{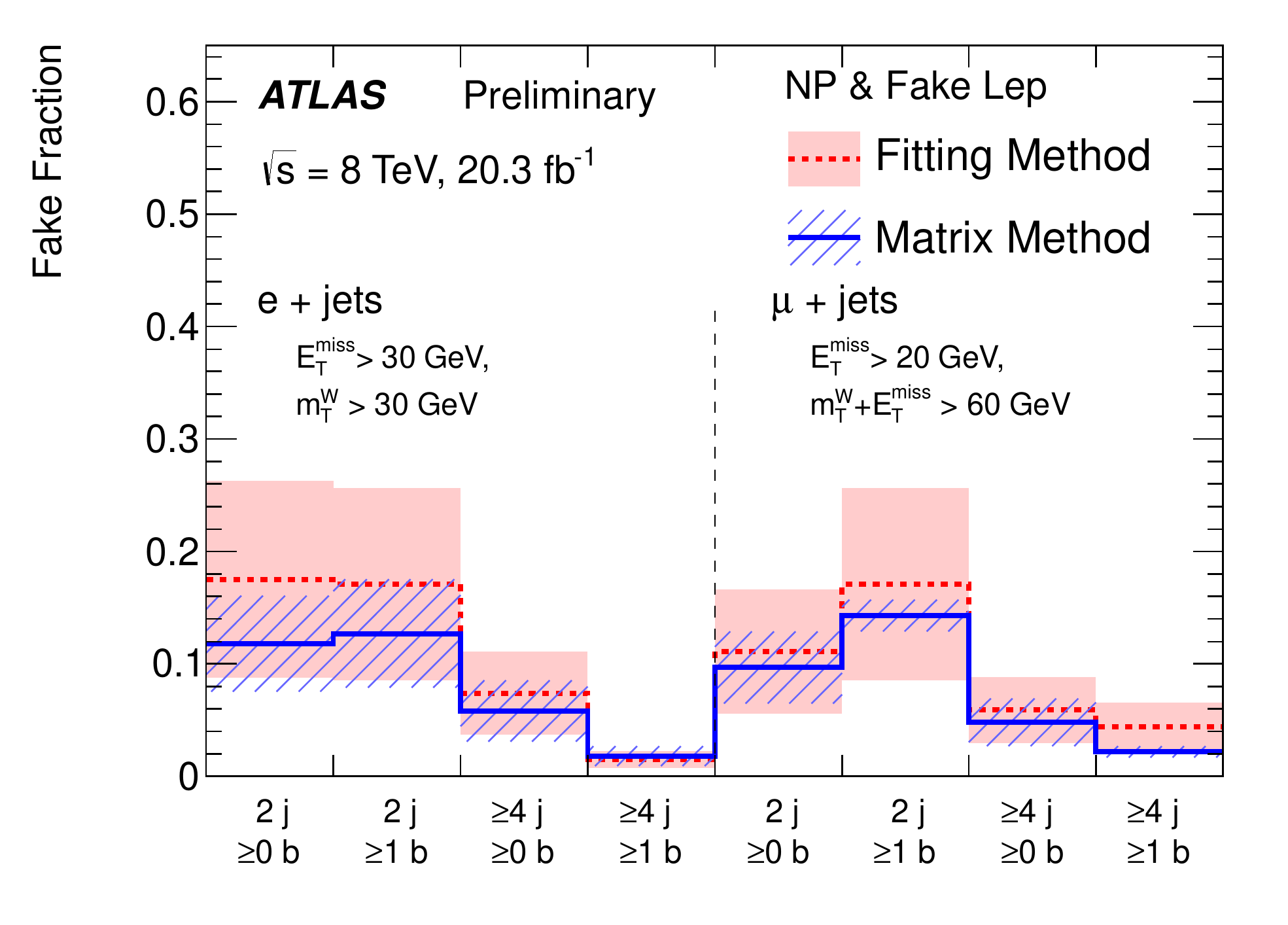}
\caption{\label{fig:fakesMMFitMethod}Comparison of the non-prompt and fake lepton background
  predictions by the matrix method and the fit method divided by the
  number of observed events in each of the lepton+jets signal regions \cite{ATLASfakes}.}
\end{minipage} 
\end{figure}

Uncertainties associated to $b$-jets are in particular important for
some top quark measurements. The ATLAS strategy is to estimate this
uncertainty by comparing MC samples with different $b$ fragmentations
and $B$-hadron decays, and cross checking the results obtained using
data. CMS, for the case of 2011 data, accounts for this
uncertainty in the jet flavour uncertainty (which is
evaluated as the maximum difference of all jet responses). For 2012
data, a specific flavour uncertainty is evaluated for each jet
flavour. This leads to an uncertainty of 0.5\% for the case of
$b$-jets (to be compared to 1-2\% uncertainty obtained in ATLAS). CMS
has recently provided a first determination of the $b$-jet energy
correction using $Z+b$ events \cite{CMSbCorrection} with a precision
of around 0.5\%, being compatible with unity (see Figure
\ref{fig:bJetCorrection}).

The detail comparison of the procedures followed in both experiments has led to
the recommended categorisation of the jet energy scale uncertainty components and
ATLAS/CMS correlations documented in \cite{JESATLASCMS}.

For what concerns the jet calibration procedure followed at the
Tevatron experiments \cite{CDFJES,D0JES}, the main differences with respect to the LHC
strategy are the following: CDF corrects the jet energy to parton
level rather than to particle level. D0 does calibrate the jets to
particle level but applies an also called absolute calibration in the sense that
the same calibration procedure is applied to both data and MC. Both
Tevatron experiments do also correct for the underlying event while
this is not the case at the LHC. The D0 experiment has recently
provided a new calibration in which dedicated corrections are applied
to light, $b$ and gluon jets \cite{D0JES}. 

\section{$b$-tagging}
The four experiments have developed various algorithms to identify
$b$-jets. Those mostly employed in top quark analyses, use the fact that
due to the long lifetime of $B$-hadrons, tracks with large impact
parameters as well as displaced vertices can be found. This kind of information is
combined in different ways in the $b$-tagging algorithms used by
default in the different experiments. For instance, CMS uses a
likelihood approach in the CSV tagger \cite{CMSbtagging}, while ATLAS
and D0 use a neural network in the MV1 \cite{ATLASbtagging} and MVA
\cite{D0btagging} algorithms, respectively, to provide a discriminant.


It is essential to check at which level the simulation predicts the
shapes of the discriminants used in the taggers and, if needed,
correct the simulation via data/MC scale factors. The calibration
consists then in measuring the efficiency in data and MC.  A detail
comparison of the $b$-tagging calibration strategy used in ATLAS and
CMS has recently been performed.  The general idea is to select a
sample enriched in $b$-jets and then compute the content of $b$-jets
before and after applying the $b$-tagging requirement. One possibility
is then to use a sample of jets containing muons, which is an
independent sample to that used in a top quark analyses. However, the
question is whether the obtained scale factors are also applicable to
inclusive $b$-jets, and ATLAS and CMS do take different approaches in
this respect. In ATLAS, the difference between the scale factors
obtained for the case of jets containing and not containing muons in
$t\overline{t}$ events is measured and found to be compatible within
an uncertainty of 4\%. This uncertainty is then considered as an
additional source of systematic uncertainty, which results as being
the dominant source, not being considered in CMS on the other hand.

Another possibility to get a sample enriched in $b$-jets is to use
$t\overline{t}$ events. Measurements have been performed in leptonic decay channels
using various techniques and reaching a precision of $\sim$2\% for jets of
$p_{\rm T}$ around 100 GeV, which is significantly better than that achieved
when using an inclusive sample of jets containing muons. However,
when using this calibration in top quark measurements one should take
into account that the scale factors are obtained assuming $V_{tb}
= 1$ and that their associated systematic sources are largely in
common to those affecting the top quark measurement under
consideration, meaning that correlations should be treated correctly. 

In both experiments, a good agreement between the top quark pair based
and muon jet based calibrations is found, and therefore various
combinations of the measured scale factors are provided. 
For each uncertainty source, the size as well as the strategy followed to
estimate it in each experiment has been compared. For the purpose of
combinations, it is then recommended to provide six different
components of the $b$-tagging efficiency related uncertainties, as
documented in \cite{ATLASCMSbtagging}, one of them to be treated as
uncorrelated and the others (stemming from modelling uncertainties) as
fully correlated between experiments.  

\section{Non-prompt and fake lepton background}

Among the various methods used to estimate the non-prompt and fake
lepton backgrounds in top quark analyses, the most commonly used are
the following two, which have been studied in detail in
\cite{ATLASfakes} for the case of the data collected at
$\sqrt{s}=8$~TeV during 2012:

\begin{itemize}

\item {\bf Matrix method:} It is based on the measurement of
  the lepton efficiencies for leptons with relaxed identification
  criteria to pass the requirements used in the analysis. The
  real lepton efficiencies are measured using $Z$ boson
  decaying to electron or muon pairs using a tag and probe
  technique. For the case of electrons, an additional correction
  is needed to account for efficiency differences in top quark and $Z$
  events. The fake efficiencies are measured from control regions
  enriched in fakes, subtracting the real lepton contribution as
  estimated from simulation. The efficiencies are measured as a
  function of various variables, chosen based on the actual dependence
  observed, agreement in the control regions and ensuring a
  small correlation between them. Systematic uncertainties,
  evaluated by changing the just mentioned parametrisation, using
  different control regions and varying the contribution of real
  leptons in the fake control region, range from 10-50\% (30-50\%) for
  the one lepton (two lepton) final states. 

\item {\bf Fitting method:} It consists on defining a model to predict
  the shape of the non-prompt and fake lepton background for a given
  distribution (as for instance the transverse missing energy). Using
  templates for the other processes derived from simulation, a
  likelihood fit to data is then performed to extract the
  normalisation of the non-prompt and fake lepton
  background. Systematic uncertainties are then evaluated by using
  different variables in the fit, varying the fit constraints, and
  from uncertainties in the modelling of $W/Z$+jets, leading to an
  overall uncertainty of 50\% in the one lepton channel. 

\end{itemize}

The predictions given by the two methods have been compared and found
to agree within uncertainties, as shown in Figure
\ref{fig:fakesMMFitMethod}.

\section{Summary}
The exploitation of the full data set collected by the LHC (and
previously Tevatron) experiments have led to significant improvements
on the uncertainties associated to the objects appearing in top
quark final states. Recent progress also took place at the LHC in
comparing the strategies used in ATLAS and CMS for the jet energy
scale and $b$-tagging calibrations leading to concrete recommendations
on how to treat these uncertainties in the context of LHC
combinations. 

{\bf Acknowledgments} This note has been prepared with help from James
Keaveney.

\end{document}